\documentstyle[12pt,epsf,epsfig,openbib]{article}
\begin{document}
\vspace{1cm}
\begin{center}
{\large\bf Fixed Points and Vacuum Energy of 
Dynamically Broken Gauge Theories}\\[.4cm]       
A.~A.~Natale~\footnotemark \, and 
\footnotetext{e-mail : natale@axp.ift.unesp.br}
P.~S.~Rodrigues da Silva~\footnotemark\\[.2cm]
\footnotetext{e-mail : fedel@axp.ift.unesp.br}
Instituto de F\'{\i}sica Te\'orica,                                         
Universidade Estadual Paulista\\
Rua Pamplona, 145, 01405-900, S\~ao Paulo, SP, Brazil
\end{center}
\thispagestyle{empty}
\vspace{1cm}

\begin{abstract}
We show that if a gauge theory with dynamical symmetry breaking has non-trivial
fixed points, they will correspond to extrema of the vacuum energy.
This relationship provides a different method to determine fixed points.
\end{abstract}
\newpage              						            
\section{Introduction.}

Gauge theories without scalar bosons may undergo the process of 
dynamical symmetry breaking, where dynamical masses are generated,
and we have the phenomenon of dimensional transmutation~\cite{coleman},
{\it i.e.} we basically do not have arbitrary parameters once the gauge
coupling constant $(g)$ is specified at some renormalization 
point $(\mu)$. In these theories all the physical
parameters will depend on this particular coupling. Therefore, it would
not be surprising if the dynamical masses follow a critical behavior
totally related to the one of the coupling constant.

QED is one example of a theory that may show dynamical chiral symmetry
breaking in the strong coupling regime. It has been suggested that QED 
in four dimensions at the same time that generates fermion masses, 
could develop a non-trivial ultra-violet fixed point, and this 
possibility is still under study as
recently reviewed in Ref.~\cite{azcoiti}. This fixed point
behavior could imply that four-dimensional QED is a non-trivial theory.
In three dimensions QED also suffers from dynamical symmetry breaking,
and recently it has been pointed out that it also may have a non-trivial
infrared fixed point~\cite{aitchison}. Such fixed points are determined
as zeros of the renormalization group $\beta$ function, and, generally 
speaking, they can be attractive or repulsive. According to the idea
of dimensional transmutation we can think about how this critical
behavior of the coupling constant is transmitted to other calculable
physical quantities.

One of the quantities for which we have precise methods to compute in
field theory is the vacuum energy, and we could naively think
that the fixed points would appear as extrema of the vacuum energy.  
Such intuitive idea is not new. When Wilson developed the concepts
of renormalization group and critical
phenomena~\cite{wilson}, he gave an example of the renormalization group equation 
making use of an analogy in classical physics of a ball rolling on a hill. In this
example the equation of motion of the ball in the hill potential was related to the 
renormalization group equation, and the fixed point was related to a stationary point.
Therefore, it seems natural to expect a deeper relation between fixed points and extrema
of energy also in field theory. However, we have been unable to find a proof of this in the
literature, and here we will give a simple presentation of such connection.

Our demonstration will rely heavily on two field theoretical methods: the 
inversion method proposed by Fukuda as a tool to compute non-perturbative
quantities in gauge theories~\cite{fukuda}, and the calculation of the
vacuum energy as prescribed by Cornwall et al.~\cite{cornwall,
cjt,cornwall2}. We will discuss the case of a gauge theory
without scalar bosons, with an unique coupling constant $(g)$, and we
believe that a more rigorous proof can be performed including the
case of a theory involving several coupling constants.
This relationship provides a completely different method to determine
non-perturbative fixed points, and could be tested in lattice simulations of
gauge theories.

\section{Vacuum energy and the inversion method.}

Many years ago Cornwall and Norton~\cite{cornwall} emphasized that the
vacuum energy $(\Omega)$ in dynamically broken gauge
theories could be defined as a function of the dynamical mass
\begin{equation}
m_{dyn} \equiv \Sigma (p) \equiv m \; ,
\label{e1}
\end{equation}
where $\Sigma (p)$ is the fermion self-energy, and once $m$ obey
the asymptotic behavior predicted by the operator product expansion
$\Omega =\Omega(\mu,g)$ is a completely finite quantity~\cite{cornwall,cjt,
cornwall2}.

In the case of a gauge theory without bare masses, and with an
unique coupling constant $(g)$, 
$\Omega$ must satisfy a homogeneous renormalization group equation~\cite{gross}
\begin{equation}
\left( \mu \frac{\partial}{\partial \mu} + \beta (g) \frac{\partial}{\partial g}
\right) \Omega = 0 \; .	                                                
\label{e2}                                                              
\end{equation}                                                              
On the other hand, the dynamically generated masses 
can be written as $m =\mu f(g)$~\cite{gross}, 
from what follows that $\mu (\partial m / \partial \mu ) = m$ 
and, consequently,
\begin{equation}
m \frac{\partial \Omega}{\partial m} = - \beta(g) 
\frac{\partial \Omega}{\partial g} \; . 			           	
\label{e3}                               			    
\end{equation}								    
This last and simple equation will be central to our argument, because it
relates the stationary condition for the vacuum energy
$(\partial \Omega / \partial m = 0)$~\cite{cornwall,cjt,cornwall2}, to
the condition of zeros of the $\beta$ function.  

We will suppose that we have a gauge theory with a critical coupling
$g_c$ separating the symmetric and asymmetric phases. For $g<g_c$ we
do not have dynamically generated masses ($m$ and $\Omega$ are equal to
zero), and we do not expect any {\it non-trivial} fixed point. For $g>g_c$
we are in the asymmetric phase, $m$ and $\Omega$ are different from
zero, and {\it as long as} $\partial \Omega / \partial g \neq 0$, {\it what 
we will show that is true when} $g>g_c$, the extrema of energy happens
for the fixed points of the theory.
Therefore, in the following we will show that 
in a gauge theory with dynamically generated masses (or condensation),
the condition for an extrema of the vacuum energy:
\begin{equation}
\beta(g) \left. \frac{\partial \Omega}{\partial g} 
\right|_{\partial \Omega / \partial m = 0} = 0 \; ,
\label{e4}
\end{equation}
always imply $\beta(g)=0$.

Before discussing the behavior of the theory for $g>g_c$,
it is interesting to see what are the consequences if the
theory has a fixed point at $g_{\star}=g_c$. Note that the
notion of critical point for the chiral transition and fixed 
point should not necessarily coincide, although this is exactly
what is expected in QED. 
A bifurcation of the self-energy, or dynamical mass, will happens between the
symmetric and asymmetric regions, and we can perform a very simple analysis
to know the behavior of the mass (and the vacuum energy) in the neighborhood
of the fixed point $g_{\star}$. The dynamical mass is given by
\begin{equation}
m = \mu \, exp\left[ - \int^g \frac{dg}{\beta (g)} \right] \; ,
\label{e4a}
\end{equation}
and assuming that near $g_{\star}$ the $\beta$ function can be approximated by
\begin{equation}
\beta (g) = b(g - g_{\star}) + ... \; ,
\label{e4b}
\end{equation}
where $b$ is a constant, we obtain
\begin{equation}
m = \mu \, (g-g_{\star})^{- 1/b} \; .
\label{e4c}
\end{equation}
On dimensional grounds the vacuum energy will be given by $\Omega \propto
c m^4$, where $c$ is a calculable negative number~\cite{gross}. For
$g<g_{\star}$ the dynamical mass is zero (as well as the vacuum energy),
and for $g>g_{\star}$ it deviates from zero according to Eq.(\ref{e4a}).
The  dynamically broken phase exist for $g>g_{\star}$, and in this case 
$g_\star$ coincides with the critical coupling for mass generation $g_c$.
For $b>0$ we note that 
the vacuum energy as a function of the coupling constant will behave
as displayed in Fig.(1a). In Fig.(1a) $\Omega = 0$ for $g<g_{\star}$, it has a
bifurcation at $g_{\star}$, and its absolute value diminishes as $g$
is increased. For $b<0$ the vacuum energy $\Omega$ is depicted
in Fig.(1b). According to the sign of $b$ we have a local minimum or a
maximum of the vacuum energy at the point $g=g_{\star}$. The
discontinuity in Fig.(1a) is artificial, and it will depends
on our ability to compute the dynamical mass behavior at the
transition point. Actually this will be a measure of the
phase transition order. 
The above argument is valid only on the vicinity of the
fixed point, where Eq.(\ref{e4b}) is reliable.

We can now proceed to the case when
$g_\star >g_c$
(the possibility that $g_{\star}=g_c$ as was discussed above,
is nothing else than a limit of this case).
Our demonstration will not depend on the specific form of the
$\beta$ function.
Considering that in the broken phase $m \neq 0$ and that the 
functional derivative of $\Omega$ with respect to $m(\Sigma )$ vanishes
at the extrema of the vacuum energy (or when $m$ is a solution of
the Schwinger-Dyson equations)~\cite{cornwall,cjt,cornwall2}, we can assert
that the extrema of energy will occur at the fixed points just showing
that in the right-hand side of Eq.(\ref{e3}),
$\partial \Omega / \partial g \neq 0$ for $g>g_c$.

To compute the vacuum energy in dynamically broken gauge theories we
need to introduce a bilocal field source $J(x,y)$, since 
we are interested in theories which admit
condensation of composite operators as, for instance, $\left\langle
\bar{\psi} \psi \right\rangle$~\cite{cjt}. $\Omega$ will be calculated after
a series of steps starting from the
generating functional $Z(J)$~\cite{cjt}:
\begin{eqnarray}
Z(J) &=& \exp [\imath W(J)]= \int d\phi 
\exp \left[ \imath \left( \int d^4 x {\cal L}(x) \right. \right.\nonumber \\ 
&& + \left. \left. \int d^4x d^4y {\phi}(x)J(x,y){\phi}(y) \right) \right] \; ,     	
\label{e5}                                                              
\end{eqnarray}                                                           
where $\phi$ can be a fermion or gauge boson field.
From the generating functional we determine the effective action $\Gamma(G)$
which is a Legendre transform of $W(J)$: 
\begin{equation}                  
\Gamma(G)=W(J)- \int d^4x d^4y G(x,y)J(x,y) \; ,                                            
\label{e6}							            
\end{equation}                                                              
where $G$ is a complete propagator, and from Eq.(\ref{e6}) we obtain
\begin{equation}
\delta \Gamma / \delta G(x,y)=-J(x,y) \; . 
\label{e7}
\end{equation}
The physical 
solutions will correspond to $J(x,y)=0$, which will reproduce the 
Schwinger-Dyson equations (SDE) of the theory~\cite{cjt}.
In general, if $J$ is the source of the operator $\cal{O}$, we have~\cite{gkm}
\begin{equation}    
\left. \frac{\delta \Gamma}{\delta J } \right|_{J=0} 
= \left\langle 0 \left| \cal{O} \right| 0 \right\rangle \; .
\label{e8}
\end{equation}
For translationally invariant (t.i.) field configurations
we can work with the effective potential given by                              
\begin{equation}                      
V(G) \int d^4x = - \Gamma(G)|_{t.i.} \; .
\label{e9}
\end{equation}
Finally, from the above equations we can define the vacuum energy as~\cite{cjt,
cornwall2}
\begin{equation}
\Omega = V(G) - V_{pert}(G) \; ,                                                       
\label{e10}							            
\end{equation}                                                              
where we are subtracting from $V(G)$ its perturbative counterpart, and
$\Omega$ is computed as a function of the nonperturbative propagators
$G$, {\it i.e.} its self-energies, $\Sigma$ or $\Pi$, whether we are working with
fermions or gauge bosons, and is zero in the absence of mass
generation. Ultimately, $\Omega$ is a function of the
dynamical masses of the theory. Note that in the cases when we do
not have bare masses $V_{pert}(G)=0$, and $\Omega = V(G)$~\cite{cjt,cornwall2}. 

There is a long discussion in the literature if the vacuum energy
$\Omega$ can be identified with the effective potential as 
described above, if the effective potential is single-valued, gauge-invariant,
etc...~\cite{banks}. 
However, we stress that all these
problems are absent at the stationary points of the vacuum energy~\cite{sico},
where even different formulations of the effective potential for composite
operators lead to the same stationary point~\cite{hay,inoue}, and it is
exactly for these points that we must compute Eq.(\ref{e4}).
We can now write Eq.(\ref{e4}) in the following form:
\begin{equation}
 - \beta(g) \left[ \frac{\partial \Omega}{\partial J}
\frac{\partial J}{\partial g} \right]_{J=0} = 0 \; .
\label{e11}
\end{equation}
However, $\partial \Omega / \partial J = - \partial \Gamma /
\partial J$, and as a consequence of Eq.(\ref{e8}) we have
\begin{equation}
\beta(g) \left\langle 0 \left| \cal{O} \right| 0 \right\rangle
\left. \frac{\partial J}{\partial g} \right|_{J=0} = 0 \; .
\label{e12}
\end{equation} 
The vacuum condensate $\left\langle 0 \left| \cal{O} \right| 0 \right\rangle$
is different from zero for $g>g_c$. Therefore,
it remains to show that ${\partial J / \partial g}|_{J=0}$
is also different from zero in the same condition, what can be
accomplished through the so called ``inversion method''~\cite{fukuda}.

Fukuda has devised a very ingenious method to determine nonperturbative
quantities~\cite{fukuda}. He noticed that to compute a nonperturbative
quantity like $\left\langle 0 \left| \cal{O} \right| 0 \right\rangle 
\equiv \vartheta$, the usual procedure is to introduce a source $J$
and to calculate the series:
\begin{equation}
\vartheta = \sum_{n=0}^{\infty}  g^n h_n(J) \; .
\label{e13}
\end{equation}
In practice we have to truncate Eq.(\ref{e13}) at some finite order,
and it gives us only the perturbative solution $\vartheta =0$ when 
we set $J=0$. The right-hand side of Eq.(\ref{e13}) should be
double valued at $J=0$ for another solution to exist, which is
not the present case. The alternative method proposed by
Fukuda is to invert Eq.(\ref{e13}), solving it in favor of $J$
and regarding $\vartheta$ as a quantity of the order of unity.
We obtain the following series:
\begin{equation}
J=  \sum_{n=0}^{\infty} g^n k_n (\vartheta) \; ,
\label{e14}								    
\end{equation}	                      					    
where the $k_n$'s satisfying $n \leq m$ ($m$ being some finite integer)
are calculable from $h_n$ also satisfying $n \leq m$. We can find
a nonperturbative solution of $\vartheta$ by setting $J=0$
through a truncated version of Eq.(\ref{e14}). The details of
the method can be found in Ref.~\cite{fukuda}. The important
point for us is that by construction of Eq.(\ref{e14}) we
verify that when $J=0$ and $\vartheta \neq 0$ ({\it i.e.}
$g>g_c$), the same value of $\vartheta$
that satisfy Eq.(\ref{e14}) leads trivially to
\begin{equation}
\partial J / \partial g |_{J=0} \neq 0 \; .
\label{e14a}
\end{equation}
Therefore, the two terms, $\partial J / \partial g$ and
$\left\langle 0 \left| \cal{O} \right| 0 \right\rangle$, of Eq.(\ref{e12})
{\it never can be equal to zero in the broken phase}!
According to this and looking at Eq.(\ref{e12}), the only
possibility to obtain $\partial \Omega / \partial m = 0$ (for $g>g_c$) 
is when we have a
fixed point ($\beta (g) = 0$), from where comes our main assertion that fixed points
are extrema of the vacuum energy. It is also clear that if $\beta (g) \neq 0$
in the condensed phase we are forced to say that the critical
and fixed point coincide ($g_c = g_\star $), and, of course,  it corresponds to an
extreme of energy. 

\section{Examples and summary.}

We would like to show some specific calculations of the
above expressions to see how they can be computed in practice.
As an example that the terms $\partial \Omega / \partial J$ and 
$\partial J / \partial g$ of Eq.(\ref{e11}) are different
from zero in the condensed phase, we can compute them in the case of
four-dimensional quantum electrodynamics ($QED_4$), 
which is supposed to have a non-trivial 
ultra-violet fixed point~\cite{azcoiti}.
In $QED_4$ the term $\partial \Omega / \partial J$ will be given by $\left\langle
\bar{\psi} \psi \right\rangle$, {\it i.e.} we have dynamical mass generation
and a condensate is formed when the gauge coupling constant $\alpha \equiv e^2/ 4\pi$
is larger than a certain critical value $\alpha_c$.
The term $\partial J / \partial g$ is also different from zero for
$\alpha > \alpha_c$, and these terms can be computed with the
help of the inversion formula determined in Ref.~\cite{ukita}.
We will not repeat here all the steps of Ref.~\cite{ukita},
where the inversion method was applied to $QED_4$, and Eq.(\ref{e14})
was obtained up to two-loop level in a gauge invariant way,
whose result is:
\begin{eqnarray}
J &=& \frac{4 \pi^2}{N_f \Lambda_f^2} \left[ 1 - \frac{\alpha}{\alpha_c} \right] 
\left\langle \bar{\psi} \psi \right\rangle \nonumber \\ 
&& + \frac{64 \pi^6}{\eta N_f^3 \Lambda_f^8} \frac{\alpha}{\alpha_c}
\left\langle \bar{\psi} \psi \right\rangle^3 \ln^2 
\left( \frac{16\pi^4}{N_f^2\Lambda_f^4 \Lambda_p^2}\left\langle \bar{\psi} 
\psi \right\rangle^2 \right) \nonumber \\ 
&& + {\cal O}(\left\langle \bar{\psi} \psi \right\rangle^3
\ln\left\langle \bar{\psi} \psi \right\rangle^2),   	       				    
\label{e15}								    
\end{eqnarray}								    
where $\alpha_c=2\pi/3 \eta$,
$N_f$ is the number of flavors, 
$\eta=\Lambda_p^2/\Lambda_f^2$, the $\Lambda 's$
are ultraviolet cutoffs associated to the photon and fermion 
self-energies~\cite{ukita}, and in the following we will assume
that $\Lambda_f=\Lambda_p=\Lambda$. The solution
$J=0$ exists only for $\alpha > \alpha_c$, and the expression of 
$\left\langle \bar{\psi} \psi \right\rangle$ for this range of coupling constant
can be easily obtained. For instance, near $\alpha =
\alpha_c$ it behaves as
\begin{equation}
\left\langle \bar{\psi} \psi \right\rangle \cong - \frac{\Lambda^3 N_f}
{8 \pi^2} \frac{(1-\alpha_c/\alpha)^{1/2}}{\ln((1-\alpha_c/\alpha)^{1/2}/2)} \; . 
\label{e16}
\end{equation}
Substituting the result of $\left\langle \bar{\psi} \psi \right\rangle$
into $\partial J / \partial g$ we verify that this term is different from zero
for $\alpha > \alpha_c$. In the case of $QED_4$ Eq.(~\ref{e11}) will have a zero only
if there is a fixed point for $\alpha \geq \alpha_c$. Note that for
$\alpha < \alpha_c$ there is not mass generation and $\Omega
\equiv 0$, and at $\alpha = \alpha_c$ we may expect a bifurcation
in the vacuum energy due to the phase transition. It is obvious
in this case that the critical and fixed points are the same.

As a final example we would like to present a calculation of
the vacuum energy showing its behavior with respect to the coupling
constant. Actually, a few years ago one of us~\cite{natale} computed 
$\left\langle \Omega \right\rangle$, which denotes the values of
$\Omega$ at the stationary points~\cite{castorina}, in the case of quenched 
$QED_4$. $\left\langle \Omega \right\rangle$
was computed using approximate solutions of the Schwinger-Dyson
equations for the fermion propagator, and the minimum of energy
was obtained for each value of the coupling constant ($\alpha$).
It was observed that the deepest minimum occurs exactly for the
critical value of the coupling constant expected to be a fixed
point. This calculation was also extended to include the effect of
four-fermion interactions~\cite{montero}. Here we will compute
$\left\langle \Omega \right\rangle$ in the case of QED taking
into account the vacuum polarization effects as discussed by
the authors of Ref.~\cite{guz,oli}. Note that this is not going
to be a proof that QED in this approximation has a fixed point, because
we will not include the effect of the four-fermion interactions
(which can change completely the result),
and we will make use of very rough approximations to the 
Schwinger-Dyson equations for the dynamical mass. A complete
calculation of $\left\langle \Omega \right\rangle$ without
the approximations that we are going to make it is not an
easy task, and could only be performed numerically.
However, we do find a behavior similar to
the one presented in Fig.(1), indicating that such studies
have still to be pursued. $\left\langle \Omega \right\rangle$
is given by the following expression~\cite{castorina}:
\begin{equation}
\left\langle \Omega \right\rangle = 2 \imath N_f
\int \frac{d^4 p}{(2\pi )^4}
\left[ \ln [1-\Sigma^2(p)/p^2]+\Sigma^2(p)/[p^2-\Sigma^2(p)] \right] \; ,
\label{e17}
\end{equation}
where $\Sigma (p)$ is the fermion self-energy of QED. Considering the
photon polarization $\Pi (p^2)=(\alpha N_f /3\pi )\ln (\Lambda^2/p^2)$
and defining
\begin{equation}
z = \frac{3\pi}{\alpha N_f} + \ln \frac{\Lambda^2}{p^2} \; , 
\label{e18}
\end{equation}
the fermion self-energy in the large $z$ limit has the form
(in the Landau gauge)~\cite{guz,oli}
\begin{equation}
\Sigma (p^2)=z^\gamma [C_1\Phi (a,c;z)+C_2 \Psi (a,c;z) ] \; ,
\label{e19}
\end{equation}
where $\gamma =3/(2\sqrt{N_f} )$, $a=\gamma (1-\gamma)$, $c=1+2\gamma$,
and $\Phi (a,c;z)$, $\Psi (a,c;z)$ are the confluent hypergeometric
functions.

Eq.(\ref{e19}), for a given number of flavors, exist only above a
certain critical coupling constant. These critical couplings have
been determined numerically, and up to four fermion flavors they
are~\cite{guz,oli,kondo}: $\alpha_c =2.00 (N_f=1), \, 2.75(N_f=2),
\, 3.51(N_f=3), 4.31(N_f=4)$. These couplings would be candidates
for fixed points as in quenched QED. Eq.(\ref{e17}) can be computed
numerically with the use of Eq.(\ref{e19}). A quite reasonable
approximation to the full result can also be obtained noticing
that the infrared behavior of Eq.(\ref{e19}) at leading
order is weakly dependent on the coupling constant~\cite{oli}.
Therefore we can expand Eq.(\ref{e17}) for small values of
$\Sigma/p$, obtaining
\begin{equation}
\frac{8\pi^2}{m^4N_f}\left\langle \Omega \right\rangle \approx
- \frac{1}{2} \int_1^{(\Lambda/m)^2} \; dx \,
\frac{\overline{\Sigma}^4}{x} + {\cal O}
\left( \frac{\overline{\Sigma}^6}{x^2} \right) \; ,
\label{e20}
\end{equation}
where $\overline{\Sigma} = \Sigma/m$ and $x=p^2/m^2$. In agreement with the
approximation leading to Eq.(\ref{e20}), $\left\langle \Omega \right\rangle$
can be calculated with the help of the asymptotic forms of Eq.(\ref{e19}),
for which (when $\Lambda/m \gg 1$) we have~\cite{guz,oli}
\begin{equation}
C_1 \approx - \frac{m^3}{\Lambda^2} e^{-\frac{3\pi}{\alpha N_f}}
\left[ \frac{3\pi}{\alpha N_f} + \ln{\frac{\Lambda^2}{m^2}} \right]^{\gamma^2}
\; , \; C_2 \approx m 
\left[ \frac{3\pi}{\alpha N_f} + \ln{\frac{\Lambda^2}{m^2}} \right]^{-\gamma^2}
\; ,
\label{e21}
\end{equation}
and
\begin{equation}
\Phi (a,c;z)_{ \; z \rightarrow + \infty }^{ \;\;\;\;\ \sim } \;
\frac{\Gamma (c)}{\Gamma (a)} e^{z} z^{(a-c)} \; ,
\; \Psi (a,c;z)_{\; z \rightarrow + \infty }^{ \;\;\;\;\ \sim } \; z^{-a} \; .
\label{e22}
\end{equation}
As noticed in Ref.~\cite{oli} only at the momentum scale of ${\cal O}(m)$ the
solution $C_1\Phi (a,c;z)$ becomes significant compared to
$C_2\Psi (a,c;z)$. The use of these asymptotic expressions is
reliable in a considerably large region of momenta, since the full
numerical solution of the Schwinger-Dyson equation can be fitted
by $C_2\Psi$ in almost all the interval of the integral of Eq.(\ref{e20})
in the case of $N_f=1$~\cite{oli}. However, the solution given by
Eq.(\ref{e19}) becomes a poor approximation for the numerical solution
of the full self-energy equation when $N_f \geq 2$~\cite{guz}.
To compute the vacuum energy we used the asymptotic expressions
given by Eq.(\ref{e22}), and the values of $\left\langle \Omega \right\rangle$
are not expected to match the ones that would be obtained with the 
complete solution of the self-energy as we increase $N_f$, but they
will keep roughly the same behavior, due to the fact that the
integral in Eq.(\ref{e20}) depends essentially on the ultraviolet
behavior of $\Sigma (p)$. We assumed $(\Lambda/m)^2 = 10^{15}$, and
kept the infrared cutoff equal to $100$ in order to be consistent
with the approximation of small values of $\Sigma /p$ that lead us
to Eq.(\ref{e20}). 

Our results for the calculation of the vacuum
energy at stationary points are shown in Fig.(2), where
$\left\langle \Omega \right\rangle$ is plotted as a function of
$\alpha$ for $N_f = 1,2,$ and $3$. As in the case of quenched QED
(without~\cite{natale} and with four-fermion interaction~\cite{montero}),
the points of minimum energy occur exactly at the critical
couplings $\alpha_c = 2.00,\, 2.75$ and $3.51$ respectively, which
are the ones expected to be fixed points. Again, we stress
that the discontinuity reflects only the poor handling of
the dynamical mass at the phase transition point. With this
calculation we could say that QED when considering vacuum polarization
effects may have non-trivial fixed points. Unfortunately,
there are indications that the introduction of four-fermion
interactions leads to a non-interacting theory~\cite{rakow}.
It would be interesting
to compute $\left\langle \Omega \right\rangle$ taking into account
the effect of four-fermion interactions (with coupling constant
$G$), and the full solution
of the fermionic self-energy with the vacuum polarization effects. 
In this case $\left\langle \Omega \right\rangle$ would appear
as a surface in the space of the coupling constants ($g,G$). We
can predict the following possibilities: a) A non-trivial fixed
point would be indicated by a minimum of the vacuum energy for finite
values of the coupling constants. b) A trivial theory would come
out with a minimum located at infinity for a infinite value of
the four-fermion interaction as observed in Ref.~\cite{rakow}. 
Although this calculation cannot be performed analytically it may provide
useful information on the strong coupling regime of QED. 

The connection between fixed points and vacuum energy
provides a totally different way to find fixed points.
As $\left\langle \Omega \right\rangle$ is a gauge invariant
physical quantity it can be computed in numerical lattice
simulations of gauge theories for different 
values of the coupling constant.
Therefore, it is possible to obtain the curve of minima
of energy in the regions of small and strong coupling,
approaching the region of the phase transition,
and according to our previous discussion the point of
minimum energy connecting these different regions will
indicate the fixed point.

In conclusion, we have shown that the extrema of the vacuum
energy are associated to the fixed points of dynamically
broken gauge theories. The vacuum energy is identically
zero for $g<g_c$, where $g_c$ is the coupling separating
the symmetric and broken phases. For $g>g_c$ we are in
the broken phase, and,
using the ``inversion method'', it 
is possible to verify explicitly that the extrema condition
(Eq.(\ref{e11})) can have a zero, only if the $\beta$ function has a zero.
In the expected case that $g_{\star}=g_c$, i.e. critical and fixed
points coincide, we use a very
simple argument to show that it is an extremum of energy. This
case can be seen as a limiting one when these points necessarily
do not coincide. Our demonstration can be extended to
the cases with several coupling constants, and possibly may be established
on more formal grounds.
We also presented an example of vacuum energy calculation which
shows the expected behavior discussed here.
It would be interesting if such relationship could
be investigated by other methods, such as
direct lattice calculations.

\section*{Acknowledgments}                    

We have benefited from discussions with O. Eboli, J. Montero 
and V. Pleitez. This research was supported in part by the
Conselho Nacional de Desenvolvimento
Cientifico e Tecnologico (CNPq)(AAN), and by Funda\c c\~ao
de Amparo a Pesquisa do Estado de S\~ao Paulo (FAPESP) (PSRS).


\newpage
\section*{Figure Captions}

\noindent

{\bf Fig.(1)} Expected behavior of the vacuum energy in a theory with
a fixed point at $g_{\star}=g_c$. a) For  positive values of $b$;
b) For negative values of $b$. 

{\bf Fig.(2)} The vacuum energy at stationary points 
$\left\langle \Omega \right\rangle$ calculated as a function of the coupling
constant $\alpha$, in the case of QED with $N_f$ fermions and considering
effects of vacuum polarization. The curves are for $N_f = 1,2,3$, and
the minima of energy are respectively at $\alpha = 2.00, \, 2.75$ and
$3.51$.
\newpage
\begin{figure}[htb]
\epsfxsize=.94\textwidth
\begin{center}
\leavevmode
\epsfbox{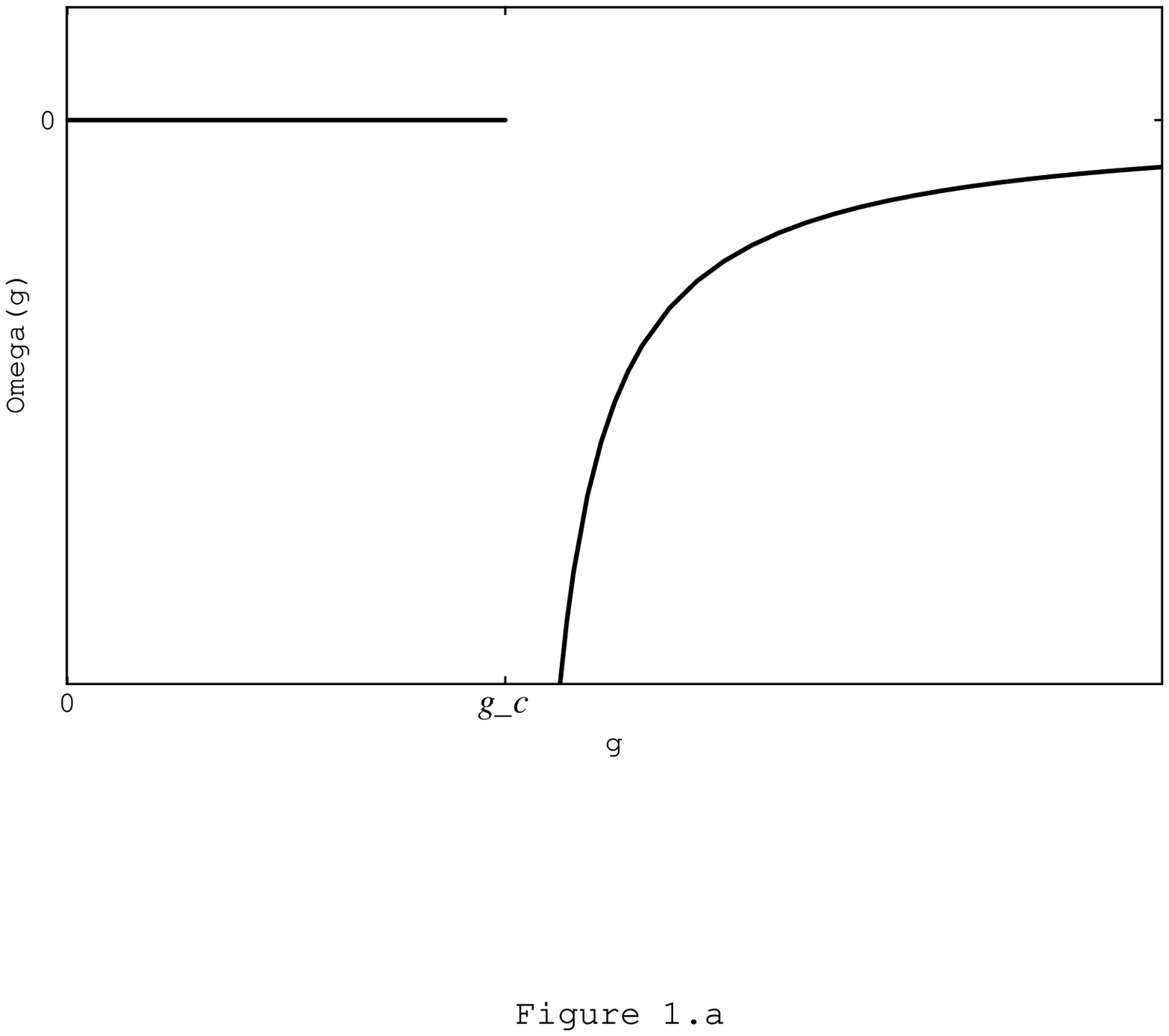}
\end{center}
\end{figure}
\begin{figure}[htb]
\epsfxsize=.94\textwidth
\begin{center}
\leavevmode
\epsfbox{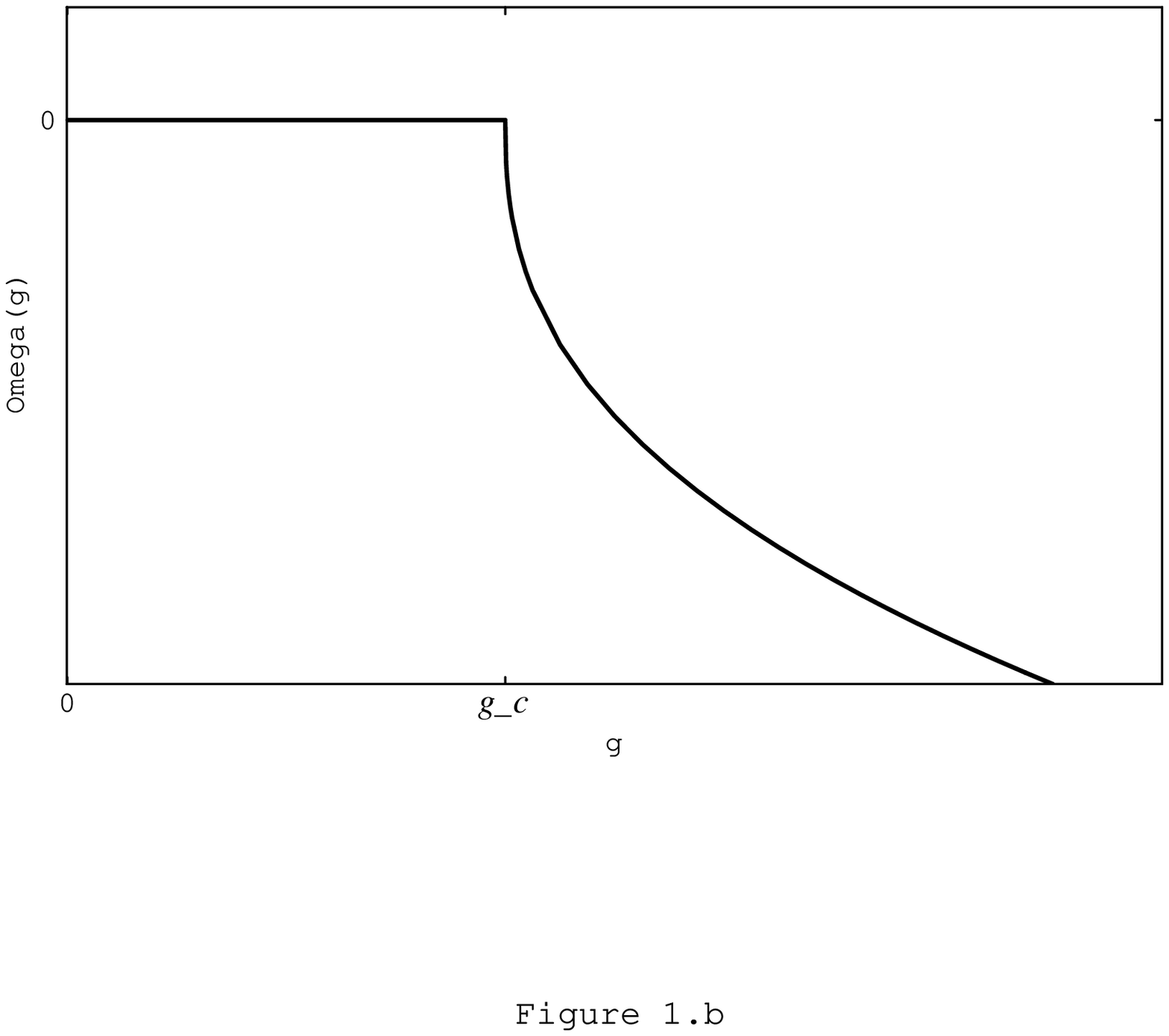}
\end{center}
\label{fig:1}
\end{figure}
\begin{figure}[]
\epsfxsize=0.9\textwidth
\begin{center}
\leavevmode
\epsfig{file=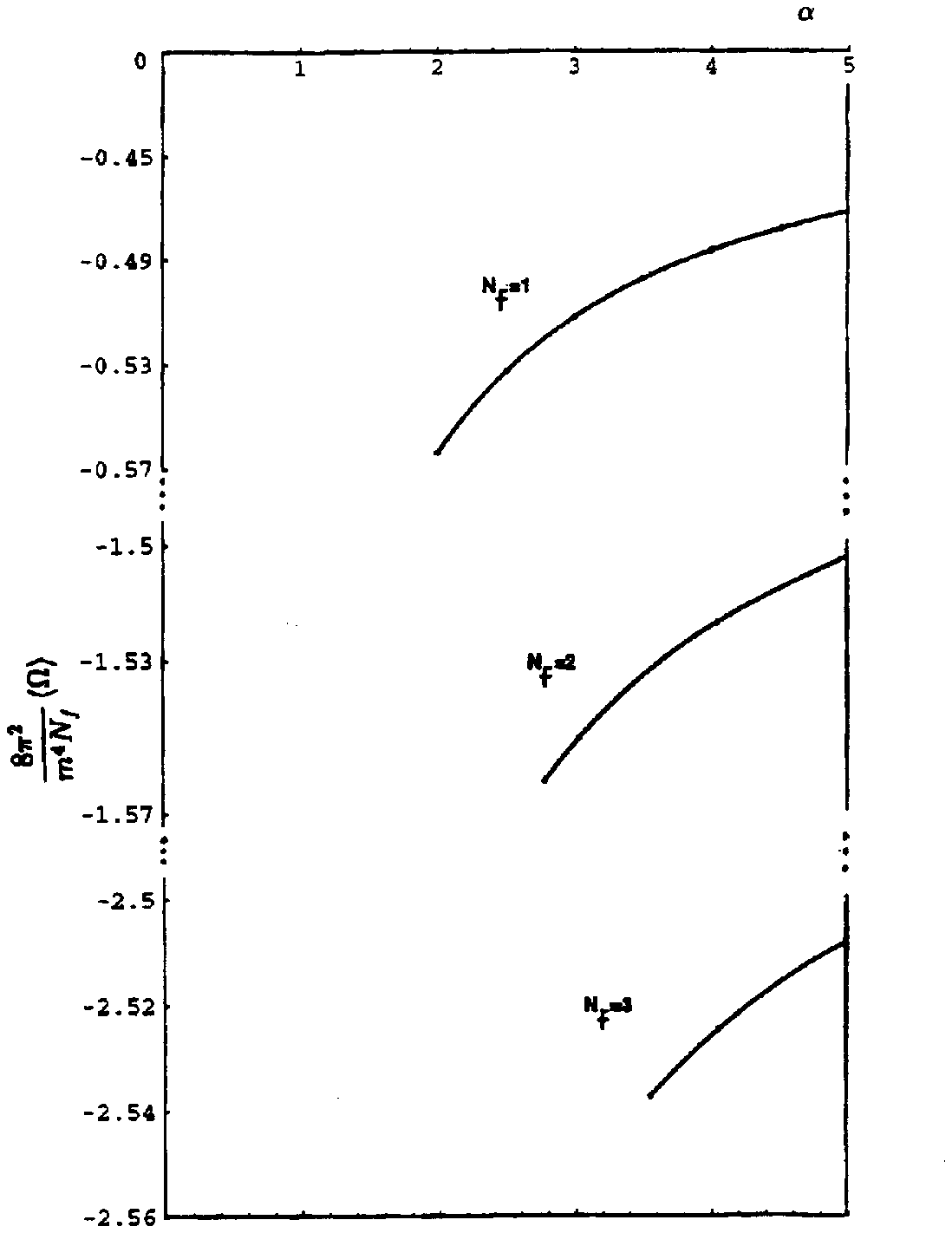,angle=180}
\vskip 1cm
\bf{Figure 2}
\end{center}
\label{fig:3}
\end{figure}
\end{document}